\documentstyle[aps,prl,epsf,floats]{revtex} 
\draft

\begin{document}
\twocolumn[\hsize\textwidth\columnwidth\hsize\csname
@twocolumnfalse\endcsname 

%\begin{flushright}
%UCLA/99/TEP/24
%\end{flushright}

\preprint{UCLA/99/TEP/24} 

\title{Unstable superheavy relic particles as a source of neutrinos
responsible for the ultrahigh-energy cosmic rays}

\author{Graciela Gelmini$^1$ and Alexander Kusenko$^{1,2}$}
\address{$^1$Department of Physics and Astronomy, UCLA, Los Angeles, CA
90095-1547 \\ $^2$RIKEN BNL Research Center, Brookhaven National
Laboratory, Upton, NY 11973}

\date{August, 1999}

\maketitle
             
\begin{abstract}
Decays of superheavy relic particles may produce extremely energetic
neutrinos.  Their annihilations on the relic neutrinos can be the origin of
the cosmic rays with energies beyond the Greisen-Zatsepin-Kuzmin cutoff.
The red shift acts as a cosmological filter selecting the sources at some
particular value $z_e \pm \delta z$, for which the present neutrino energy
is close to the $Z$ pole of the annihilation cross section.  
We predict no directional correlation of the ultrahigh-energy cosmic rays
with the galactic halo.  At the same time, there can be some directional
correlations in the data, reflecting the distribution of matter at red
shift $z=z_e \pm \delta z$.  Both of these features are manifest in the
existing data.  Our scenario is consistent with the neutrino mass reported
by Super-Kamiokande and requires no lepton asymmetry or clustering of the
background neutrinos.

\end{abstract}

\pacs{PACS numbers: 98.70.Sa, 95.85.Ry, 14.60.Pq, 95.35.+d \hspace{1.0cm}
UCLA/99/TEP/24} 

\vskip2.0pc]

%%%%%%%%%%%%%%%%%%%%%%%%%%%%%%%%%%%%%%%%%%%%%%%%%%%%%%%%%%%%%%%%%%%%%%
%% Section I %%%%%%%%%%%%%%%%%%%%%%%%%%%%%%%%%%%%%%%%%%%%%%%%%%%%%%%%%
%%%%%%%%%%%%%%%%%%%%%%%%%%%%%%%%%%%%%%%%%%%%%%%%%%%%%%%%%%%%%%%%%%%%%%

\renewcommand{\thefootnote}{\arabic{footnote}}
\setcounter{footnote}{0}

%\section{Introduction}
%\label{sec-1}

Cosmic rays~\cite{data} with energies beyond the Greisen-Zatsepin-Kuzmin
(GZK) cutoff~\cite{gzk} present a challenging outstanding puzzle in
astroparticle physics and cosmology.  One of the most appealing and
economical explanations, proposed by T.~Weiler~\cite{weiler}, is based on
the observation that the GZK cutoff is absent for
neutrinos~\cite{weiler,fargion}.  High-energy neutrinos can annihilate near
to Earth on the background relic neutrinos via the $Z$
resonance~\cite{weiler} producing the ultrahigh-energy cosmic rays (UHECR).
A very appealing feature of this scenario is that the requisite energy
scale appears naturally~\cite{gk} if the relic neutrinos have mass $m_\nu =
m_{_{SK}}\equiv 0.07^{ +0.02}_{-0.04}$~eV, as inferred from the
Super-Kamiokande (SK) data (assuming no degeneracy in the neutrino mass
spectrum).  Such relic neutrinos can efficiently annihilate a neutrino with
energy $E_\nu = M_{_Z}^2/2 m_{_{SK}} \sim 10^{23}$~eV, corresponding to a
$Z$ resonance.  This energy happens to be just a few orders of magnitude
above the GZK cutoff, which is precisely what is needed to explain the GZK
puzzle.

The only problematic element in this scenario~\cite{weiler,gk} so far is 
the source of the ultrahigh-energy neutrinos.  In this paper we show that
heavy unstable relic particles can produce the requisite neutrinos.  The
spectrum of these neutrinos is different from that of known astrophysical
neutrino sources (it grows rapidly with energy, up to a sharp cutoff), so
that the upper bounds of Refs.~\cite{wb} do not apply.  The flux of
ultrahigh-energy neutrinos may be high enough to explain UHECR regardless
of the lepton asymmetry.  

Heavy relic particles decaying into hadrons and photons have been
considered as sources of UHECR~\cite{particles,kt1,sigl,kt}.  However, the
lack of directional correlation of the observed events with the galactic
halo and the observed clustering of events~\cite{takeda} in the directions
unrelated to the distribution of the galactic dark matter may stymie this
explanation.  Indeed, if the decaying particles are the origin of the
ultrahigh-energy protons and photons, one expects the directions of UHECR
to point to some nearby sources.  The present data are too
sparse to reach a definite conclusion~\cite{anisotropy}.

If, however, the superheavy relic particles decay into neutrinos, the
directional correlation with the local distribution of matter is not
expected in general.  The only significant contribution to the
$\nu\bar{\nu}$ annihilation ($\nu\bar{\nu} \rightarrow Z \rightarrow p
\gamma ...$) on background neutrinos with mass $m_\nu$ comes from 
energies near the $Z$ pole, so the high-energy neutrino must have energy
close to $E_{\rm res} = M_{_Z}^2/2 m_{\nu}$ to produce an appreciable
amount of cosmic rays.  Let us suppose that the ultrahigh-energy neutrinos
come from a two-body decay of some superheavy particle $X$ into a neutrino
and some other light particle.  If the mass of the relic particle $m_{_X}$
is not fine-tuned to be exactly $2 E_{\rm res}$, the local distribution of
$X$-particles has no effect on the highest-energy cosmic rays.  If
$m_{_X}>2 E_{\rm res}$, the neutrinos produced at red sift $z_e= (m_{_X}/2
E_{\rm res}-1)$ annihilate very efficiently close to Earth because their
present energy corresponds to the $Z$ pole.  The red shift acts as a
cosmological filter selecting the sources at some particular value $z_e$.
The directions of the UHECR should, therefore, point to sources at $z=z_e$,
not to those in the local group.  One may still observe clustering of
events congenial to the distribution of luminous matter at $z \approx z_e$,
but not at $z \approx 0$.

Let us consider some superheavy relic particle $X$ that can presently
contribute to the cold dark matter~\cite{kt1,wimpzillas}.  Such particles
could, for example, be produced non-thermally at preheating~\cite{kt}.  Let
us also assume that it has a lifetime $\tau_{_X}$ that is much greater than
the age of the universe $t_0$ and that it decays with emission of at least
one neutrino with branching ratio $B_\nu$.  We assume that hadronic and
other visible decay modes have a negligible branching ratio.  For
simplicity, we will consider a two-body decay with emission of a neutrino
with energy $E_{\nu,e} = m_{_X}/2$.  Generalizations to many-body processes
and other modes of decay are straightforward. 

The spectrum of ultrahigh-energy neutrinos from decaying unstable heavy
particles is given by~\cite{stecker,ggs}
\begin{equation}
E \frac{d\phi}{dE} = \frac{3}{2} \phi_{\gamma,0} Y_{_{X,0}} B_\nu 
{t_0 \over \tau_{_X}}
\left (\frac{E}{m_{_X}/2} \right )^{3/2} \theta({m_{_X} \over 2}-E), 
\label{spectrum}
\end{equation}
where we have assumed $\tau_{_X} \gg t_0$.  Here $\phi_{\gamma,0} =
n_{\gamma,0}/4\pi \approx 10^{12} {\rm cm}^{-2} {\rm s}^{-1} {\rm sr}^{-1}$
is the flux of the cosmic microwave background photons, and $Y_{_{X,0}}$ is
the ratio of the $X$ number density to that of photons, $Y_{_{X,0}}=
n_{_{X,0}}/n_{\gamma,0}$.  We assume that the universe is matter-dominated
from the time of emission until present. As usual, index $0$ refers to the
present time, while subscript $e$ denotes the time of emission.

We will now compute the flux of UHECR from the annihilation of
ultrahigh-energy neutrinos on the background neutrinos with mass $m_\nu$.
The cross section for the process $\nu \bar{\nu} \rightarrow Z^*
\rightarrow {\em hadrons} $ has a sharp maximum at the $Z$ pole: $\sigma
\propto 1/ \left \{ (s-M_{_Z}^2)^2 + M_{_Z}^2 \Gamma_{_Z}^2 \right \}$,
where $s=2 E_{\nu} m_\nu$, and $\Gamma_{_Z}\approx 2.49$~GeV is the $Z$
width. We are interested in the neutrino annihilations within the
attenuation length of protons and photons, which is about 50~Mpc from
Earth.  Within this radius one can neglect the red shift.  Only those
neutrinos that have energies $E_{\nu,0} =E_{\rm res} \pm \delta E =
(M_{_Z}^2/2 m_\nu) [1\pm  \Gamma_{_Z}/M_{_Z}] $, can contribute to the flux
of UHECR.  The flux $\Delta \phi$ of these neutrinos can be obtained by
integrating $d\phi(E)/dE$ in equation (\ref{spectrum}) from $E_{\rm res}
-\delta E$ to $E_{\rm res} +\delta E$:
\begin{equation} 
\Delta \phi = 3 \phi_{\gamma,0} Y_{_{X,0}} B_\nu {t_0 \over \tau_{_X}}\left 
(\frac{E_{\rm res}}{m_{_X}/2} \right )^{3/2}
\left ( \frac{\Gamma_{_Z} }{M_{_Z}} 
\right ) . 
\label{fraction} 
\end{equation}

The value of $n_{_{X,0}}=\Omega_{_X} \rho_c/m_{_X}$ is bounded
from above by the requirement that $\Omega_{_X} h^2 < 0.25 $.  $X$
particles can be the cold dark matter~\cite{kt1,wimpzillas} if $\Omega_{_X}
h^2 \approx 0.1-0.25$.

The probability $P$ for each of these neutrinos to interact inside the GZK
sphere, that is within 50~Mpc from Earth, depends on the lepton asymmetry
parameter $\eta =n_{\nu,{\rm relic}} /n_{\gamma,0}$.  We treat $\eta$ as a
parameter whose value is constrained by nucleosynthesis, as
well as  large-scale structure and cosmic microwave background (CMB)
radiation anisotropy measurements.  

If the universe has zero lepton asymmetry, $\eta=0.14$.  However, depending
on the cosmological parameters, this may not be the value that gives the
best agreement with observations.  If the cosmological constant is
small or zero, $\Omega_\Lambda < 0.1 $, the value $\eta \approx 4$ is
favored by the combination of the large scale structure and the CMB
data~\cite{as,lp}.  For $\Omega_\Lambda \approx 0.5 $, the range $0< \eta
<2$ is allowed~\cite{lp}.  For larger values of the cosmological constant,
$\Omega_\Lambda> 0.7$, lepton asymmetry must be small~\cite{as,lp,kr}, and
$\eta \approx 0.14$.

The $\nu\bar{\nu}$ annihilation mean free path for energies above the $Z$
pole is much greater than the Hubble distance $H^{-1}$.  Therefore, those
neutrinos whose present energy is $E_{\rm res} \pm \delta E$ travel
unabsorbed to red shifts $z_a = 2 \Gamma_{_Z}/M_{_Z}=0.06$, which
corresponds to distance $280 \, {\rm Mpc} (0.65/h)$.  Within this distance
the mean free path is $\lambda = 1/(\sigma_{{\rm ann},_Z} n_{\nu,{\rm
relic}})= 5.3 \eta^{-1} \times 10^{28}$~cm, where $\sigma_{{\rm ann},_Z} =
4\pi G_{_F}/\sqrt{2}$.  Thus, within 50~Mpc from the observer a fraction 
\begin{equation}
P = 3.8 \times 10^{-4} \left ( {\eta \over 0.14} \right )
\end{equation}
of the ultrahigh-energy neutrinos annihilate and produce ``$Z$-bursts''. 
The nucleons and photons from $Z$ decays undergo a cascade of scatterings 
off the background photons but their energies remain largely above the GZK
cutoff~\cite{gk}.  The resulting spectrum of these UHECR will be discussed
in detail in an upcoming paper~\cite{gknv}.  Taking into account the
attenuation and absorption, each annihilation event
produces $N \approx 10$ photons and protons~\cite{weiler} whose energies
exceed the GZK cutoff.  The flux of UHECR beyond the GZK cutoff is
predicted to be
\begin{eqnarray}
\phi_{_{CR}} & = &  
{1.7  \over  
(4\pi \, {\rm sr}) \ {\rm km}^2 \ (100 \, {\rm yr})} 
\nonumber \\ &  \times &
\left ( \frac{N}{10} \right )  
\left (\frac{\eta}{0.14} 
\right )
\left (\frac{\Omega_{_X}}{0.2}
\right ) 
\left (\frac{h}{0.65}
\right )^2 \nonumber \\ &  \times &
\left (B_\nu \frac{10^7 t_0}{\tau_{_X}}
\right )
\left (\frac{0.07 \, {\rm eV }}{m_\nu}
\right )^{3/ 2}
\left (\frac{10^{14} \, {\rm GeV}}{m_{_X}}
\right )^{5 / 2} . 
%\end{array}
\label{CRflux} 
\end{eqnarray}

The mass of the $X$ particle must exceed twice the resonance energy $E_{\rm
res}= M_{_Z}^2/2 m_\nu = 5.9 (0.07 {\rm eV}/m_\nu)\times 10^{13}$~GeV.
According to Ref.~\cite{ggs}, in the mass range $m_{_X}
\stackrel{>}{_{\scriptstyle \sim}} 10^{12}$~GeV, heavy particles decaying
into neutrinos are allowed by the present data as long as 
\begin{eqnarray}
{\tau_{_X} \over t_0} B_\nu^{-1} & > & 2.4 \times 10^5 
\nonumber \\ &  \times &
\left ( {\Omega_{_X} \over 0.2} \right ) 
\left ( {h \over 0.65} \right )^2 \left ( 
{10^{14} \, {\rm GeV} \over m_{_X}} 
\right)^{3/4}.
\label{tauX}
\end{eqnarray}

As pointed out by G.~Sigl (private communication), the limit set by the
Energetic Gamma Ray Experimnent Telescope (EGRET) on the diffuse low-energy
gamma-ray flux resulting from the $Z$-bursts could constrain this model.
This question deserves further study.  The other bounds usually mentioned
in connection with the ultrahigh-energy neutrinos~\cite{wb} do not apply
because the spectrum of neutrinos in equation (\ref{spectrum}) is peaked
near $m_{_X}$.  This differs drastically from the inverse power-law
spectrum assumed in Refs.~\cite{wb} for astrophysical sources.  In our
case, the number of low-energy neutrinos is negligible, and the only upper
bound on the neutrino flux comes from the observations of atmospheric
showers~\cite{ggs}.

The flux of super-GZK cosmic rays in equation (\ref{CRflux}) is consistent
with the data for some range of parameters around $\tau_{_X} \sim
(10^5-10^7) t_0$, $m_{_X} \sim (10^{13}-10^{15})$~GeV, $m_\nu \sim (0.01 -
1.0) $~eV.  It is particularly interesting that the UHECR puzzle can be
explained without fine-tuning the masses and the lifetimes of the
hypothetical $X$ particles.  The mass $m_{_X}$ has to be at the scale where
one generally expects to find some new physics.  The same weakly
interacting $X$ particles could be the cold dark matter~\cite{wimpzillas}. 
Another interesting feature of our scenario is that the neutrino mass can
be in the SK range.  If the neutrino density is enhanced by either
gravitational clustering~\cite{weiler} (for neutrino masses $\sim$~eV) or
due to a lepton asymmetry of the universe~\cite{gk}, the parameter space
for $m_{_X}$, $\tau_{_X}$ increases accordingly.

Since there is no natural reason for $m_{_X}$ to be exactly equal to $2
E_{\rm res}$, the neutrinos produced by the $X$ particle decays in the
local cluster of galaxies give a negligible contribution to the measured
flux of cosmic rays.  However, those particles that decayed at red shift
\begin{equation}
z_e={m_{_X} \over 2 E_{\rm res}} -1 = \frac{m_{_X} m_\nu}{M_{_Z}^2} -1 
\label{redshift}
\end{equation}
have just the right energy at present for an efficient annihilation.  The
arrival directions of UHECR point, therefore, to their sources located in a
thin spherical shell at red shift $z_e\pm \delta z$, where $\delta z=
(1+z_e) (\Gamma_{_Z}/M_{_Z})=0.03 (1+z_e) $.

If $m_{_X}\sim 10^{14}$~GeV for $m_\nu=m_{_{SK}}$,
then $z_e\sim 0.1-1$.  In this case, one may hope to identify the
sources with the remote clusters of galaxies at red shift $z_e$.  In
principle, this would allow one to measure $m_{_X}$.  The present data
already shows some directional clustering of events~\cite{takeda}.  There
is a $1\%$ probability for these correlations to be
accidental~\cite{takeda}.  In the future, their statistical significance
can be tested on a larger data sample.  Recent studies of anisotropies in
the UHECR~\cite{anisotropy} concentrated on the dark matter distribution in
our galaxy and the local group of galaxies.  In view of our results, a
similar analysis of possible correlations with the known clusters of
galaxies at some fixed (but yet unknown) red shift could reveal the
locations of distant sources and help determine $m_{_X}$.

In any case, our scenario does not predict any correlation of the
directions of UHECR with the distribution of local dark matter.  If the X
particle decays are many-body, $\delta z$ is increased because of a wider
spectrum of the emitted high-energy neutrinos. 

Non-thermal production of superheavy relic particles at the end of
inflation~\cite{kt1,wimpzillas} and their role in
cosmology~\cite{heavycosm} have been the subject of intense studies
recently.  It is clear that $\Omega_{_X}$ is a model-dependent parameter,
which can be $\sim 0.1$ in many realistic models~\cite{kt}.

We have presented a plausible candidate source of ultrahigh-energy
neutrinos that can be the explanation of the cosmic rays beyond the GZK
cutoff.  Weakly interacting massive particle with mass $m_{_X}
\stackrel{>}{_{\scriptstyle \sim}} 10^{13}$~GeV and lifetime $\tau_{_X}
\stackrel{>}{_{\scriptstyle \sim}} 10^5 t_0$ can decay into very energetic
neutrinos.  The expansion of the universe red shifts the energies of the
neutrinos produced in a spherical shell at $z = (m_{_X} m_\nu /M_{_Z}^2 -1)
$ to the value for which the cross section of $\nu \bar{\nu}$ annihilation
near Earth is large.  These annihilations can create a flux of photons and
nucleons with energies above the GZK cutoff that is sufficient to explain
the present data.  The possible directional correlations of UHECR events
produced in this manner reflect the matter distribution at red shift $z_e
\pm \delta z$.  Our scenario is consistent with the neutrino masses in the
range reported by the Super-Kamiokande experiment.

This work was supported in part by the US Department of Energy grant
DE-FG03-91ER40662, Task C.

%%%%%%%%%%%%%%%%%%%%%%%%%%%%%%%%%%%%%%%%%%%%%%%%%%%%%%%%%%%%%%%%%%%%%%
%% References %%%%%%%%%%%%%%%%%%%%%%%%%%%%%%%%%%%%%%%%%%%%%%%%%%%%%%%%
%%%%%%%%%%%%%%%%%%%%%%%%%%%%%%%%%%%%%%%%%%%%%%%%%%%%%%%%%%%%%%%%%%%%%%

\end{document}